\documentclass[journal]{IEEEtran}
\usepackage{cite}
\usepackage{graphicx}
\usepackage{comment}
\usepackage{url}
\usepackage{listings}
\usepackage{algorithmic}
\usepackage{array}
\usepackage{fixltx2e}
\usepackage{stfloats}
\usepackage{ifpdf}
\usepackage{amsmath}
\usepackage{tcolorbox}
\usepackage{caption}
\makeatletter
\@ifpackageloaded{xcolor}{%
  \PassOptionsToPackage{table}{xcolor}%
}{}
\makeatother
\usepackage[table]{xcolor}

\usepackage{soul}
\sethlcolor{yellow}
\usepackage{amsmath}
\usepackage{colortbl}
\begin{document}
\title{Towards Sustainable Computing: \\
Exploring Energy Consumption Efficiency of Alternative Configurations and Workloads in an Open Source Messaging System}

\author{
    \IEEEauthorblockN{
        Maria Voreakou$^*$, 
        George Kousiouris,
        Mara Nikolaidou
        }
        
    \IEEEauthorblockA{
        Harokopio University of Athens, Athens, Greece\\
        \{voreakou, gkousiou, mara\}@hua.gr
    }
    \thanks{ Part of the research leading to the results presented in this paper has received funding from the European Union’s funded Project HUMAINE under grant agreement No. 101120218. }
}




\maketitle

\begin{abstract}
Energy consumption in current large scale computing infrastructures is becoming a critical issue, especially with the growing demand for centralized systems such as cloud environments. With the advancement of microservice architectures and the Internet of Things, messaging systems have become an integral and mainstream part of modern computing infrastructures, carrying out significant workload in a majority of applications. In this paper, we describe an experimental process to explore energy-based benchmarking for RabbitMQ, one of the main open source messaging frameworks. The involved system is described, as well as required components, and setup scenarios, involving different workloads and configurations among the tests as well as messaging system use cases. Alternative architectures are investigated and compared from an energy consumption point of view, for different message rates and consumer numbers. Differences in architectural selection have been quantified and can lead to up to 31\% reduction in power consumption. The resulting  dataset is made publicly available and can thus prove helpful for architectures' comparison, energy-based cost modeling, and beyond. 
\end{abstract}

\begin{IEEEkeywords}
Sustainable Computing, Messaging Systems, RabbitMQ, Energy Consumption, Testbed, Sustainable Architectures, Open Energy Dataset
\end{IEEEkeywords}

\IEEEpeerreviewmaketitle

\section{Introduction}

\IEEEPARstart{}Sustainability and Green computing \cite{greenComputing} are important topics of the current era, with multiple and complex applications across large data centers, consuming more and more energy as an integral part of our digital life. Data centers' energy consumption has started to become an issue in recent years \cite{energyImpact} \cite{metrics}.
Therefore, careful consideration needs to be applied on the design, configuration, and operation of systems and software under various loads, in order to reduce energy consumption, by closely monitoring and analyzing different strategies, optimizations, and understanding the behavior of such systems.

Messaging systems have emerged as one of the main backbones of modern application and system architectures. Messaging systems act as a main backbone interconnecting different systems, in a System of Systems approach, in order to enforce greater separation, enhance scalability and fault tolerance, as well as enable abstraction and functional separation between systems. They are extensively used either as data distribution systems (e.g. in IoT platforms \cite{thingsboard}), generic load balancing mechanisms (e.g. in the case of the queue-based load leveling pattern \cite{microsoftQueueBasedLoad}), as core cloud services (e.g. Amazon SNS/SQS) or as the main mechanisms in social networks \cite{socialNetworks}. Therefore, they are highly relevant in a distributed, large-scale, and cloud native context, carrying out significant workload in day-to-day activities. They also ensure message delivery in faulty environments with out of the box functionalities for message tracking, delivery guarantees and cross-system notifications. 

The scope of this study is to extract the energy footprint of one of the most commonly used open source messaging systems, RabbitMQ, under varying incoming and outgoing loads, represented through the parameters of input messages per second, number of routing keys as well as number of consumer queues. Furthermore, we take into consideration different strategies in its configuration, such as the selection of the main type of exchange for the same usage scenario.

RabbitMQ consists of 3 different exchanges: \textit{Direct}, \textit{Topic}, and \textit{Fanout} Exchange, which define the way the messages are distributed to the consumers through the queues. \textit{Direct} creates a simple Exchange with a strict filtering pattern on the message, so the consumers subscribed to this exchange receive all the messages sent to this queue that exactly match the desired routing key. Similarly, \textit{Topic} creates an exchange with a partial or full match filtering pattern, and the consumers subscribed to the specific topic can consume the messages from the bound queues, if the partial or full match is successful. \textit{Fanout} on the other hand is an exchange which broadcasts the message to all the queues that are bound to this exchange, so all the consumers listening to any queue of this exchange will consume all the messages.

\begin{figure}
    \centering
    \includegraphics[width=1\linewidth]{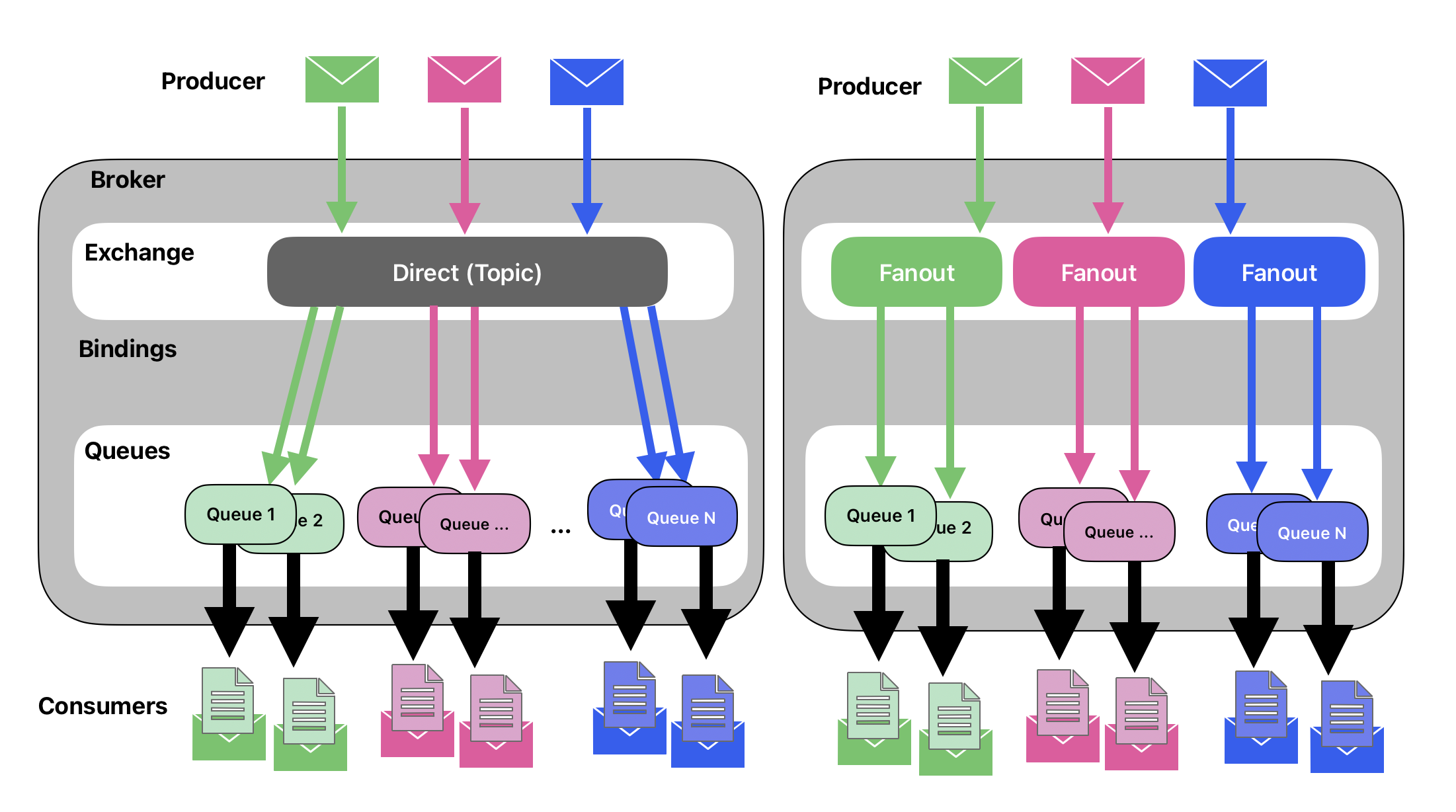}
    \caption{RabbitMQ Direct vs Fanout Exchange Experiment Strategy}
    \label{fig:rabbitmq-strategy}
\end{figure}

The goals of this experiment are as follows:
\begin{itemize}
    \item Create a testbed architecture that includes the main  components (load generator, RabbitMQ setup, energy monitoring, etc.) and define scenarios to explore, by setting parameters such as message rate, Queues, Routing keys, etc.    
    \item Map the different experimentation parameters on the energy needed for the operation of the system. The analysis of such data may result in understanding how these parameters affect the total energy consumed. Having that information may result in future specific energy-based billing models for messaging systems offered as a service, i.e. in cloud-based environments.
    \item Quantify the effect of choosing between two different strategies in order to distribute messages based on routing keys, as shown in Figure \ref{fig:rabbitmq-strategy}. While the typical strategy within RabbitMQ is to use one topic-based exchange with full key matching or a direct exchange, an alternative design may be considered in which each key is represented by a separate fanout exchange. In this case, all interested consumers are bound to each of the desired exchanges. With this, we achieve eliminating routing key matching, and thus we could potentially see different behavior in terms of energy consumption. As a trade-off, we need to have a more complex registration process and potentially larger memory needs due to the multiple exchanges' design. 
\end{itemize}

Based on the overall results and the relevant conclusions, a systems engineer should be able to define the best possible setups so that the application created is both more sustainable and better performing.

The paper is structured as follows. In Section 2, the relevant works are mentioned, in terms of relevant tools and analysis. In Section 3, we present the Project Testbed with the selected tools, including hardware and software setup, the test application and the experiment scenarios we are going to run. In Section 4, we present an initial analysis of the results including the main observed patterns and trends. Finally, in Section 5, we conclude the analysis and define future steps.

\section{Related Work}

Authors of \cite{decentralisedmqttPatients} built a decentralized MQTT for a real-time remote healthcare monitoring system using local edge devices, and measured the energy consumption using the \textit{turbostat} tool, to determine the overall system energy footprint. No special information regarding the RabbitMQ energy consumption is mentioned. However, they concluded that RabbitMQ has proved resilient in managing data during network disruptions, ensuring uninterrupted patient monitoring.

Authors of \cite{lin2024bridging} made an analysis in the current carbon estimation methodologies and proposed a usage and allocation based carbon model. Although this model aligns only with serverless computing, they concluded that it is valuable to  design models based on carbon-aware pricing which is influenced directly by the power consumption. To enforce such approaches on messaging systems, the authors proposed that relevant benchmark data such as the ones presented in this work are needed.

Authors of \cite{cadorel2024protocol} evaluated existing energy models (\textit{PowerAPI} \cite{powerapi} and \textit{Scaphandre} \cite{scaphandre}) in servers using different performance settings such as hyperthreading\cite{intelHyperThreading} and turboboost\cite{intelTurboboost} to measure the overall system usage while processing different applications.

Authors of paper \cite{RANScaphandre} evaluated their Radio Access Networks (RANs) using \textit{Scaphandre} \cite{scaphandre} and \textit{Kepler} \cite{kepler} by measuring energy metrics, in order to build power-aware scheduling algorithms.

In \cite{strempel2021measuring}, the author motivated by the ever-increasing energy consumption in data centers, measured the power consumption in applications written in C, by using \textit{Scaphandre} \cite{scaphandre} to compare different algorithms and their power consumption.

Authors of \cite{comparisoncpugpu} presented an overall evaluation of tools used to measure energy consumption including Power Profiling Software, Energy Measurement Software, as well as Energy Calculators. For real-time process-level measurements, Power Profiling software such as \textit{Scaphandre} \cite{scaphandre} is  able to estimate more processes running in parallel and it has a less complicated and more intuitive setup process for working in a virtualized environment \cite{comparisoncpugpu}. The authors  concluded that software-based power meters based on Intel RAPL \cite{rapl} and NVIDIA NVML \cite{nvidianvml} can be used to estimate consumption at a fine granularity with low overhead.

Open Source tools such as \textit{OpenMessaging} \cite{openmessaging} provide an overall benchmark solution including the most famous MQTT brokers, however they do not include relevant metrics available for power consumption.

\section{Energy Consumption Exploration Testbed}

In order to perform the envisioned experiments, a testbed is constructed, including the target application that contains a messaging system, target hardware, load generation processes, and means for collecting the relevant energy and hardware metrics.

\subsection{Test Application}
The test application is a Cloud/Fog application \cite{bagios} \cite{GeTh} that includes an edge computing component to obtain sensor readings, and a cloud backend component for data ingestion based on RabbitMQ \cite{rabbitmq}. The application consists of 3 main entities as shown in Figure \ref{fig:final-testbed}, the \textit{Backend} which is the RabbitMQ backend, the \textit{Producers} (sensors) which are responsible to send the data to RabbitMQ, and the \textit{Consumers}. The backend also includes a MongoDB, which stores data as a part of a reliability and availability of the sensors (e.g. heartbeat of a sensor) scheme. The goal in this study is to analyze the backend part, and specifically the RabbitMQ component. 

\subsection{Hardware Setup}
The involved hardware consisted of two nodes, one for the execution of the traffic load (production and consumption clients) and one for the main backend. The separation is necessary in order to keep the acquired backend measurements as pristine as possible, since load generation and consumption is by definition a very intensive process that can lead to interference.
The Hardware specifications of the node responsible to run the backend are: CPU Intel Core i7-4790K (up to 4.4GHz), 32 GB of DDR3 1600MHz RAM, 250GB Hard disk storage, running in Ubuntu OS 24.04 LTS.

\begin{figure}
    \centering
    \includegraphics[width=1\linewidth]{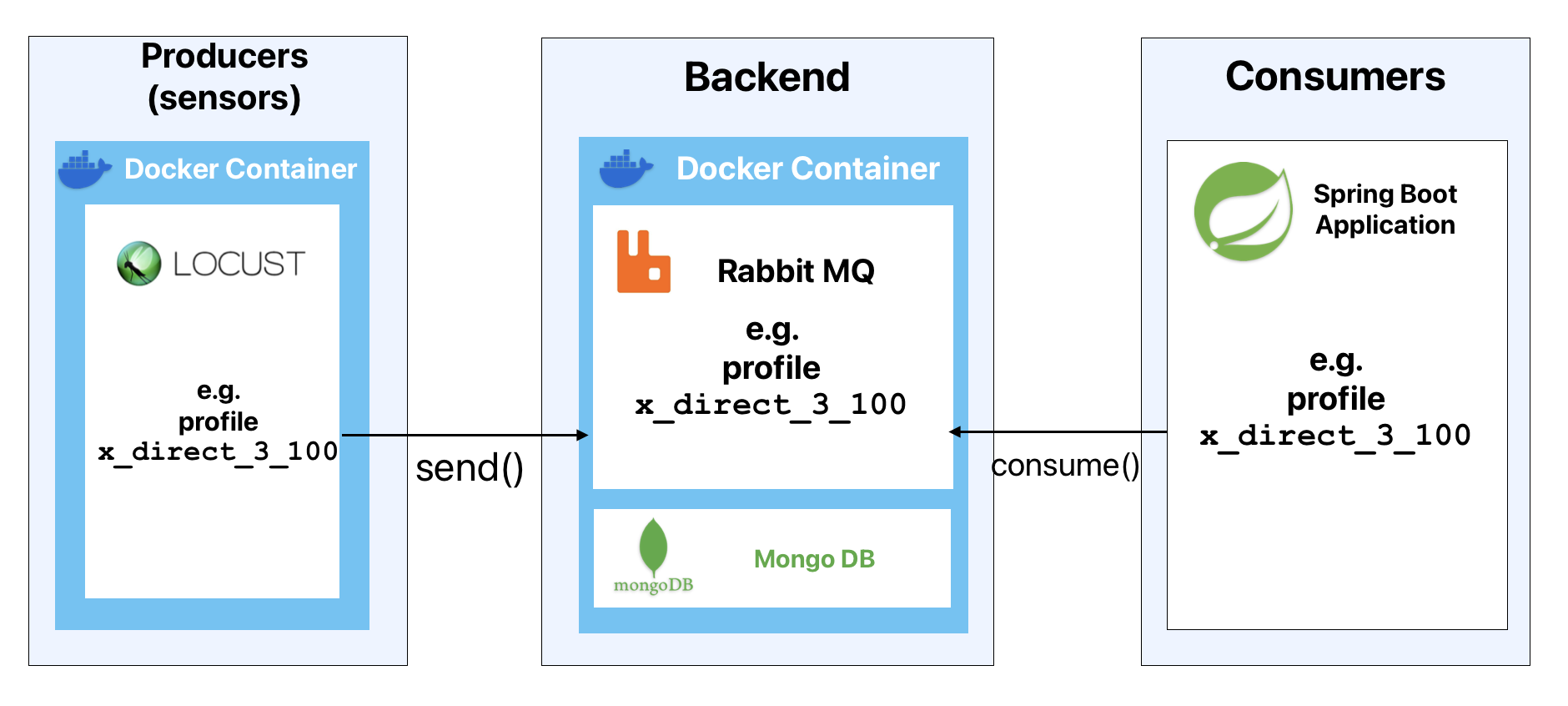}
    \caption{Testbed Overview}
    \label{fig:final-testbed}
\end{figure}

\subsection{Power Measurement Software}
To be able to measure the energy consumption on a computing system, a way is needed to get the metrics on the bare metal of the system. Libraries related with these kind of metrics are few, and some examples are \textit{Kepler} \cite{kepler}, \textit{dhsb} library \cite{dshb}, \textit{energy-tracker} library \cite{energytracker}, \textit{Open Hardware Monitor} \cite{hwmonitor}, and \textit{powerstat} \cite{powerstat}. Although the aforementioned libraries give an overview about energy consumption, they do not provide more fine grained measurements, such as power consumption per process ID.
\textit{PowerAPI} \cite{powerapi} and \textit{Scaphandre} \cite{scaphandre} are Power-Profiling software, obtaining information from CPU \& RAM directly using RAPL. RAPL stands for "Running Average Power Limit". It is a feature on Intel \& AMD x86 CPU's manufactured after 2012 which provides power measurements at a very fine-grained level \cite{scaphandre} \cite{rapl}. Both \textit{PowerAPI} and \textit{Scaphandre} are commonly used for real-time measurements. 

Both \textit{PowerAPI} and \textit{Scaphandre} are giving quite similar results compared to other existing tools \cite{cadorel2024protocol}. However, \textit{PowerAPI}'s features like Smart-watts, which is based on performance event-based regression modeling, may affect in some cases the reliability of the power consumption measurements. Furthermore,  \textit{Scaphandre}'s ability to run more parallel processes with less overhead \cite{comparisoncpugpu} made us select it as our tool of preference. Based on the model \textit{Scaphandre} uses, it calculates micro-watts per process ID. It also gives the option to run it on bare metal, or to use it with Kubernetes, and still measure metrics on bare metal. It has also been used in research to measure power consumption in data centers to reduce negative impacts such as energy waste \cite{strempel2021measuring}.


To analyze the energy and performance profile of RabbitMQ, a number of different metrics have been used, such as CPU, Memory, Disk, and traffic usage. The project's goal is to measure the energy consumption in different scenarios, getting a measured result of what are the best practices in order to achieve a more sustainable system.

In order to use \textit{Scaphandre} on our system, some configuration needs to take place. The steps taken are documented on the following GitHub repository \cite{githubVoreakou}, for validation and reproduction purposes. \textit{Scaphandre} uses a \textit{Prometheus} exporter \cite{prometheus}, and also comes dockerized together with a \textit{Grafana} dashboard \cite{grafana}, which helps to see in real-time graphs and results of the energy consumption of the application under investigation.

\subsection{Measurement Scenarios}

Measurement scenarios revolve around the main messaging system (RabbitMQ). It is reasonable to consider that energy usage may be affected by different factors such as the rate of incoming messages, the exchange type, the number of consumers, as well as how the messages get delivered from source to destination. RabbitMQ is based on the concept of routing keys. Each incoming message may be annotated by such a key that contains relevant metadata. Then a consumer can have their queue bound to an input (exchange), based on the needed metadata matching. Thus, based on the usage scenario, a given incoming message may be distributed to one or multiple consumers based on these bindings between queues and routing keys. 

In our work, we have considered two major messaging use cases. 
\begin{itemize}
    \item \textit{Account driven use case}: The scenarios consist of the same number of routing keys as the Queues (i.e. a 1:1 ratio between incoming and outgoing messages). Scenarios like this can be applicable to account based messaging where each account consumes only the messages that are specifically targeted to them, with the routing key being the user ID for example. Such systems may include e-government, e-commerce systems for ordering, banking systems etc. 
    \item \textit{Data Distribution driven use case}: The scenarios consist of a limited number of routing keys that interest multiple consumers (queues) per key. Thus an incoming message may be delivered to \textit{M} interested consumers, having a 1:M ratio between incoming and outgoing messages. This use case might be applicable to scenarios such as sensor/event data delivery in IoT platforms, Priority Messaging filtering (each key refers to different priority or type of information level), group messaging filtering (when many subscribers consume the same information),  and so on. 
\end{itemize}

For the examined use cases, and in order to implement the different design mentioned in the Introduction, we can either use a direct exchange or a topic one with a full key matching for comparing with the fanout strategy. In our case, we selected the latter so that the system is flexible to be extended in the future with partial matching use cases.

The naming structure of the project scenarios is defined based on the parameters on which we want to experiment. First, the Incoming \textbf{Message Rate} per \textbf{Second}, the \textbf{Exchange} Type this scenario relies on based on the strategy (direct/fanout) identified in Figure \ref{fig:rabbitmq-strategy}, the defined Number of \textbf{Routing Keys} (or the \textbf{Number of Exchanges} in case of a fanout scenario), as well as the Number of \textbf{Queues}. The final naming structure is:
\centerline{\textbf{MessageRate\_Second\_Exchange\_RoutingKeys\_Queues}}

All the project scenarios created for this research are mentioned in Table \ref{table_messages_consumption}. In addition to these parameters, we have also divided the scenarios under the two different use cases (\textit{Account} versus \textit{Data Distribution}) to match to realistic usages. We take as granted that the Exchanges are Durable\cite{rabbitmqAMQP091}, a RabbitMQ option that guarantees delivery and poses significant burden on the system such as increased disk I/O.

On our testbed, we have configured both Queues and Exchanges as Durable, which leads to messages to also be stored on RabbitMQ's side, in case a consumer is unable to consume messages on time.
Given this, consumers may also impact the behavior of RabbitMQ, depending on their availability and how fast they consume the messages from their queues. Messages that are in the "Ready" state (i.e. messages delivered to a consumer queue but not yet consumed by the client) are stored in the disk and memory of RabbitMQ. So a more "lazy" consumer will pose more stress on the system.

In order to include also this additional energy impact, we assumed in our implementation that consumers are not always available to receive all the messages immediately.

\subsection{Message Generation and Consumption}
In order to make the scenarios easy to run, as well as to get reliable results, various metrics across the system must be monitored. During and at the end of each experiment, we have been checking metrics related with the messages produced, message rates, RabbitMQ memory and disk usage, and metrics provided by Scaphandre, such as CPU \& Memory usage per PID, and more. Some of the metrics we find important for this paper, are indicated in Table \ref{table_messages_consumption}.

The automations and settings that were applied are:

\textit{RabbitMQ} Configurations: The following configurations were made per scenario inside the code using SpringBoot @Profiles\cite{profiles}:
    \begin{itemize}
        \item dynamic configuration of queues, bindings, and exchanges through the use of application.yml variables
        \item generate consumers using @Listener to dynamically create them
        \item adjust the Task Executors so that the mocked consumers consume in a similar rate as in real life scenarios
    \end{itemize}
    
Usage of \textit{Locust} library: Locust is an open source load testing tool \cite{locust}, used to make our load tests real-like. By using the pika library \cite{pika}, and by defining 1 worker and 1 User per 1 Task, we had the number of users be the same as the number of messages that will be published in RabbitMQ, thus acting as a publishing client.
    
Command Line Access: In order to avoid UI access overheads, which can be significant on RabbitMQ for such message rates, all accesses were terminal-based. The RabbitMQ and Grafana UIs were accessed only at the beginning and the end of each experiment.

System split in 2 nodes: The \textit{Backend} entity is installed on a different node than the  \textit{Producers} and the \textit{Consumers} of RabbitMQ messages, as shown in Figure \ref{fig:final-testbed}, in order to not interfere with the energy  results.

\section{Results}
In this section we analyze the effect of the various parameters (load rate, number of keys, number of queues) on the energy consumption of each type of exchange setup.
In order to compare the results of all the scenarios, we first need to define the values we examine, and how these will get us the conclusion for the best setup per use case. 

To be able to measure the energy consumption at a message level, but also validate the experimental process, we calculate the total messages processed per experiment, as shown in Equation \ref{eq:total_messages}, where $\textit{n}_{\textit{rate}}$ is the defined input message rate (messages/second), \textit{t} is the total time of the experiment, $\textit{n}_{\textit{queues}}$ is the number of queues and $\textit{n}_{\textit{routing keys}}$ is the number of routing keys. 

We also define the number of messages consumed as shown in Equation \ref{eq:consumed_messages}, where $\textit{n}_{\textit{em}}$ is the total messages of the experiment and $\textit{n}_{\textit{rm}}$ is the number of messages on the state \textit{Ready} in RabbitMQ at the end of the experiment. The latter indicate the finalization of their routing inside RabbitMQ and their delivery to the respective queues.

\begin{figure}
\captionsetup{labelformat=simple, labelsep=space} 
\renewcommand{\figurename}{Eq.} 
\setcounter{figure}{0} 
\begin{equation}
\text{Total Messages n}_{\textit{em} } = \textit{n}_{\textit{rate}} \cdot \textit{t} \cdot \left(\frac{\textit{n}_{\textit{queues}}}{\textit{n}_{\textit{keys}}}\right)
\end{equation}
\caption{Calculation of the total messages processed}
\label{eq:total_messages}
\end{figure}

\begin{figure}
\captionsetup{labelformat=simple, labelsep=space} 
\renewcommand{\figurename}{Eq.} 
\begin{equation}
\text{Consumed Messages} = \textit{n}_{\textit{em}} - \textit{n}_{\textit{rm}}
\end{equation}
\caption{Calculation of the messages consumed}
\label{eq:consumed_messages}
\end{figure}

\begin{figure}[t]
    \centering
    \includegraphics[width=1\linewidth]{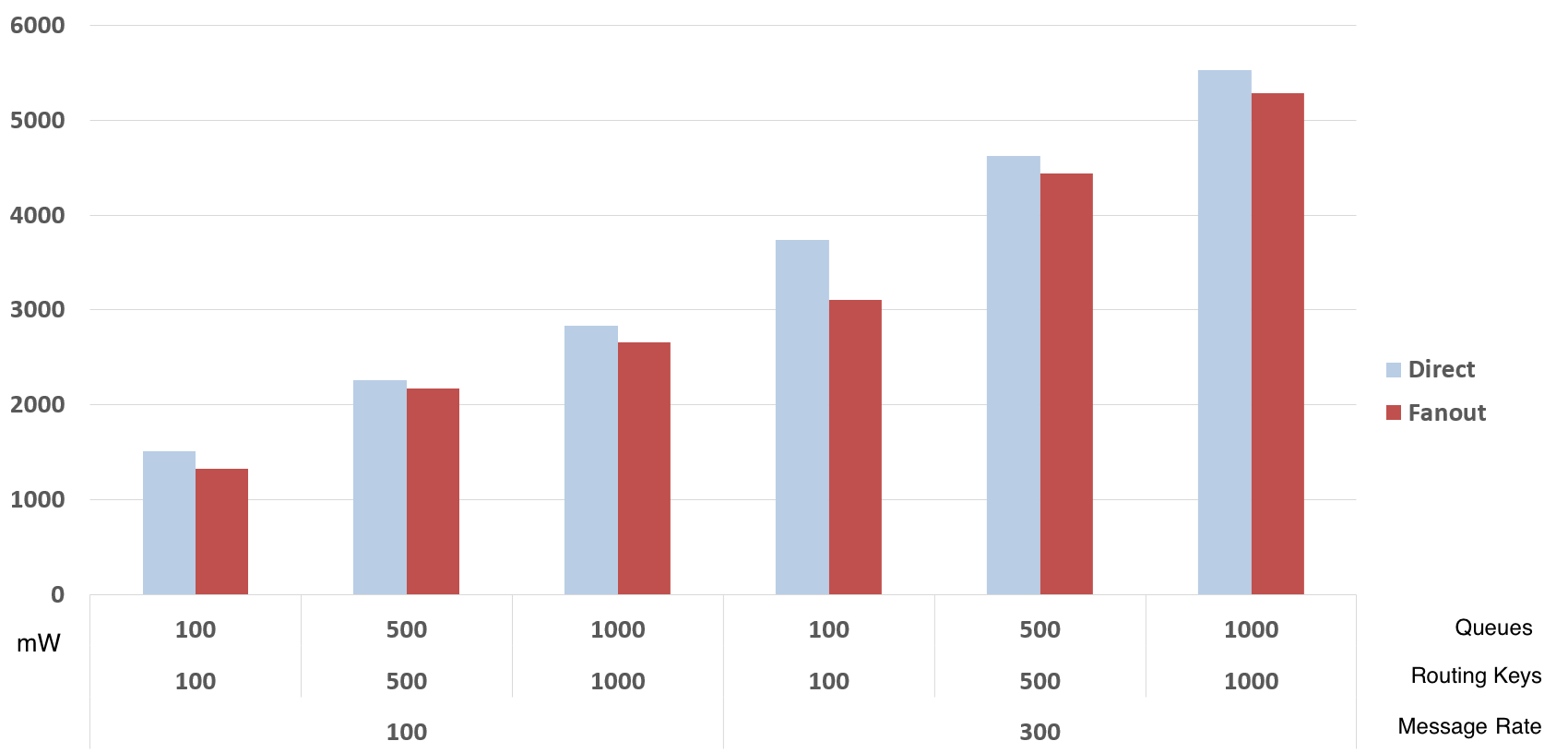}
    \caption{Account Driven Uses Cases comparison of average Power Consumption in mW}
    \label{fig:account-power}
\end{figure}

\begin{figure}[t]
    \centering
    \includegraphics[width=1\linewidth]{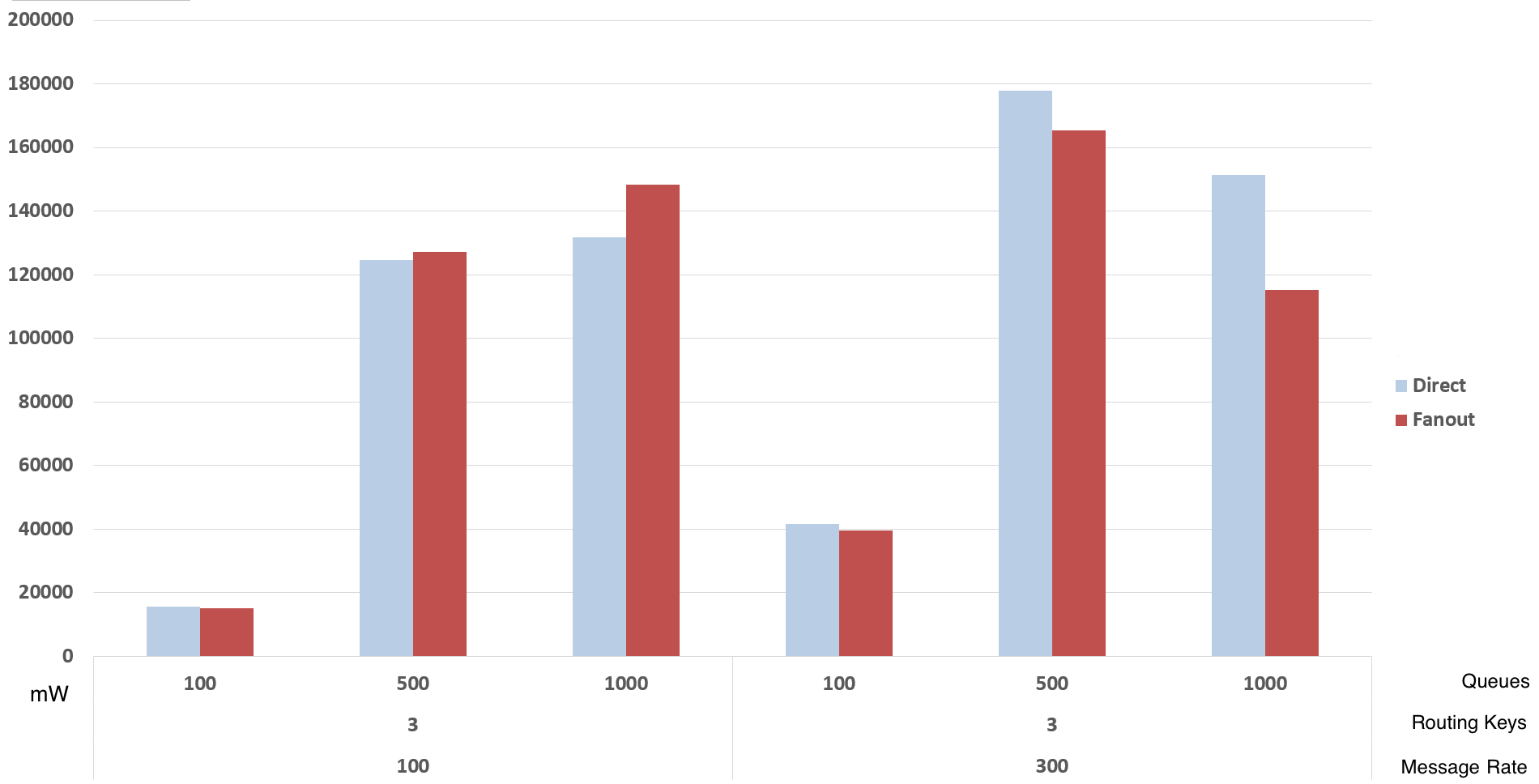}
    \caption{Data Distribution Driven Uses Cases comparison of average Power Consumption in mW}
    \label{fig:priority-power}
\end{figure}

\subsection{Analyzing the results by Use Case}
All the experiments ran in discrete and separate 30-minutes slots. A dataset was created and is accessible through a GitHub Repository \cite{githubVoreakou}, for validation and reproduction purposes. In Table \ref{table_messages_consumption}, some of the exported metrics are mentioned, in order to better describe the analysis in the rest of this Section. 

\subsubsection{Account Driven Use Case}
In Figure \ref{fig:account-power}, the results of power consumption of all the scenarios of this use case is shown. All the scenarios in this use case consist of exactly the same number of routing keys as the number of Queues, which are either 100, 500 or 1000. The Figure is also divided into 2 different message rates, one for 100 messages/second and one for 300 messages/second. In general, the results show that the \textit{Fanout} strategy had on average power consumption from 4,58\% up to 20,28\% lower than \textit{Direct}, while the \textit{Direct} always had a slightly higher power consumption.
Furthermore, we see a correlation between Energy Consumption and average Memory usage, as shown in Table \ref{table_messages_consumption}. All scenarios related with the \textit{Fanout} strategy have a lower memory usage in average, while all the \textit{Direct} scenarios have slightly increased memory usage. 
We can also see in Table \ref{table_messages_consumption}, that no Disk access (Read and Write access) is performed during any of the scenarios of this use case. This might be due to the fact that this use case's Outgoing/Incoming factor is 1, which means that each Queue is bound with just one routing key, and thus there is no additional message reproduction needed from RabbitMQ, as the message is sent to exactly one exchange.

\subsubsection{Data Distribution Driven Use Case}
In Figure \ref{fig:priority-power}, the power consumption of all the scenarios of this use case is shown. All the scenarios in this use case consist of exactly 3 routing keys, while the number of Queues is either 100, 500 or 1000. The Figure is also divided into 2 different message rates, one for 100 messages/second and one for 300 messages/second. In general, the results show that while for the scenarios with 100 messages/second, \textit{Fanout} has a slightly increased average power consumption (from 2\% up to 11,15\%), for the scenarios with 300 message/second rate the \textit{Fanout} exhibits less average power consumption (from 5\% up to 31,18\%). 
Initially this can look peculiar, however, by looking at the Table \ref{table_messages_consumption}, we see that in the scenarios with 300 messages/second rate, the \textit{Fanout} scenarios' average Disk Access and average Memory Usage is significantly lower than the \textit{Direct} scenarios' ones. For the scenarios with 100 messages/second rate, the average Disk Access and average Memory Usage look similar between the \textit{Direct} and \textit{Fanout} scenarios. The difference can affect the power consumption, and thus, the different behavior.

In some scenarios which were configured with a high incoming and outgoing message rates, as seen on Table \ref{table_messages_consumption}, RabbitMQ was not able to process the messages on time, and this resulted in the amount of messages consumed being less than expected and the memory and disk i/o to increase. Due to this stress on the messaging system, there were intermittent breaches of communication between RabbitMQ and Locust (i.e. Producers), lowering the rate of the latter. This especially happened for values 167 and 333 of the Outgoing/Incoming factor and for the cases where the incoming message rate was high (300 messages/sec). We still consider them as valid cases, as there are real scenarios for which the aforementioned issues can happen.

\subsection{Comparison of Direct versus Fanout Strategies}

Looking at all the scenarios of both use cases and comparing the \textit{Direct} with \textit{Fanout} scenarios, we see that \textit{Fanout} scenarios in most of the cases have a lower power consumption than the \textit{Direct} ones. Looking at the results on a more macroscopic level where economy of scale  plays a major role, we could possibly be looking at vast differences with significant energy impact, especially when examining a multi-node system running 24/7. 

Furthermore, looking at some specific scenarios of this experiment, we see that a small part of \textit{Fanout} scenarios have more power consumption in comparison to the \textit{Direct} cases. This means, that just by looking at the configuration, without taking into account metrics such as Disk Access and Memory usage together with Power Consumption, can lead to non optimal results.

To that end, we can conclude that in our Testbed for the \textit{Account Driven} use case, it is more energy efficient to use a \textit{Fanout} strategy instead of a \textit{Direct} strategy. For the \textit{Data Distribution Driven} use case, we see that the Input Message rate influences the power consumption significantly for the \textit{Direct} strategy, while if faced with a higher message rate, the \textit{Fanout} strategy is more efficient in terms of power consumption, again.

\begin{table*}[!t]
\captionsetup{justification=centering}
\caption{Message and Resource Usage Metrics in each of the Scenarios \\ (\textbf{Grey:} \textit{Data Distribution} Driven Use Case, \textbf{White:} \textit{Account Driven} Use Case)}
\label{table_messages_consumption}
\centering
\resizebox{\textwidth}{!}{%
\begin{tabular}{|l|l|l|l|l|l|l|l|l|}
\hline
\parbox[c]{2.5cm}{\centering \textbf{Scenario Name}} & 
\parbox[c]{1.2cm}{\centering \textbf{Expected Input Messages}} & 
\parbox[c]{1.2cm}{\centering \textbf{Actual Input Messages}} & 
\parbox[c]{1.5cm}{\centering \textbf{Expected Output Messages}} & 
\parbox[c]{1.5cm}{\centering \textbf{Actual Output Messages}} & 
\parbox[c]{1.2cm}{\centering \textbf{Outgoing/ Incoming factor}} & 
\parbox[c]{1.7cm}{\centering \textbf{Avg Power Consumption (mW)}} &
\parbox[c]{1.2cm}{\centering \textbf{Avg Disk access (MB)}} & 
\parbox[c]{1.2cm}{\centering \textbf{Avg Mem Usage (MB)}} 
\\ \hline

\rowcolor{gray!30}
100\_1\_Direct\_3\_100 & 
180000 & 
175810  & 
6000000 & 
5860333 & 
33 &
15691.76 &
0 &
248.39
\\ \hline

\rowcolor{gray!30}
100\_1\_Direct\_3\_500 & 
180000 & 
177767 & 
30000000 & 
29627833 & 
167 &
124619.42 &
135.19 &
4704.67
 \\ \hline

\rowcolor{gray!30}
100\_1\_Direct\_3\_1000 & 
180000 & 
108608 & 
60000000 & 
36202667 & 
333 &
131837.93 &
126.79 &
7806.6
 \\ \hline

\rowcolor{gray!30}
300\_1\_Direct\_3\_100 & 
540000 & 
528621 & 
18000000 & 
17620700 & 
33 &
41587.06 &
0 &
305.1
 \\ \hline

\rowcolor{gray!30}
300\_1\_Direct\_3\_500 & 
540000 & 
241746 & 
90000000 & 
40291000 & 
167 &
177771.49 &
134.77 &
7208.88
 \\ \hline

\rowcolor{gray!30}
300\_1\_Direct\_3\_1000 & 
540000 & 
118093 & 
180000000 & 
39364333 & 
333 &
151291.26 &
158.84 &
10885.45
 \\ \hline

\rowcolor{gray!30}
100\_1\_Fanout\_3\_100 & 
 180000 & 
 175650 & 
 6000000 & 
 5855000 & 
 33 &
 15098.75 &
 0 & 
 249.97
 \\ \hline

\rowcolor{gray!30}
100\_1\_Fanout\_3\_500 & 
180000 & 
175683 & 
30000000 & 
29280500 & 
167 &
127188.42 &
123.92 &
4706.97
 \\ \hline

\rowcolor{gray!30}
100\_1\_Fanout\_3\_1000 & 
180000 & 
119549 & 
60000000 & 
39849667 & 
333 &
148390.58 &
151.21 &
9384.35
 \\ \hline

\rowcolor{gray!30}
300\_1\_Fanout\_3\_100 & 
540000 & 
510377 & 
18000000 & 
17012567 & 
33 &
39603.3 &
0.01 &
297.52
 \\ \hline

\rowcolor{gray!30}
300\_1\_Fanout\_3\_500 & 
540000 & 
218325 & 
90000000 & 
36387500 & 
167 &
165369.74 &
130.29 &
6730.03
 \\ \hline

\rowcolor{gray!30}
300\_1\_Fanout\_3\_1000 & 
540000 & 
105463 & 
180000000 & 
35154333 & 
333 &
115327.5 &
83.69 &
7331.74
 \\ \hline

100\_1\_Direct\_100\_100 & 
180000 & 
177153 & 
180000 & 
177153 & 
1 &
1513.34 &
0 &
185.87
 \\ \hline

100\_1\_Direct\_500\_500 & 
180000 & 
177757 & 
180000 & 
177757 & 
1 &
2255.48 &
0 &
252.69
 \\ \hline
 
100\_1\_Direct\_1000\_1000 & 
180000 & 
177691 & 
180000 & 
177691 & 
1 &
2829.67 &
0 &
302.39
 \\ \hline

300\_1\_Direct\_100\_100 & 
540000 & 
528899 & 
540000 & 
528899 & 
1 &
3733.45 &
0 &
191.46
 \\ \hline

 300\_1\_Direct\_500\_500 & 
 540000 & 
 528665 & 
 540000 & 
 528665 & 
 1 &
 4622.04 &
 0 &
 274.09
 \\ \hline
 
300\_1\_Direct\_1000\_1000 & 
540000 & 
528830 & 
540000 & 
528830 & 
1 &
5531.35 &
0 &
369.41
 \\ \hline

100\_1\_Fanout\_100\_100 & 
180000 & 
175580 & 
180000 & 
175580 & 
1 &
1321.9 &
0 &
193.63
 \\ \hline

100\_1\_Fanout\_500\_500 & 
180000 & 
155013 & 
180000 & 
155013 & 
1 &
2169.91 &
0 &
256.57
 \\ \hline

100\_1\_Fanout\_1000\_1000 & 
180000 & 
162409 & 
180000 & 
162409 & 
1 &
2657.33 &
0 &
319.98
 \\ \hline
 
300\_1\_Fanout\_100\_100 & 
540000 & 
509578 & 
540000 & 
509578 & 
1 &
3103.79 &
0 &
192.26
 \\ \hline

300\_1\_Fanout\_500\_500 & 
540000 & 
528573  & 
540000 & 
528573 & 
1 &
4441.17 &
0.02 &
289.56
 \\ \hline

300\_1\_Fanout\_1000\_1000 & 
540000 & 
515405 & 
540000 & 
515405 & 
1 &
5289.06 &
0 &
396.46
 \\ \hline

\end{tabular}
}
\end{table*}

\section{Conclusion}
In this paper, we experimented with an existing RabbitMQ application, and we described and setup a Testbed in order to create project scenarios and benchmark our experiment strategies. The according dataset is made available to the community for further analysis.

The comparison was focused between 2 strategies: the \textit{Fanout} and the \textit{Direct} strategy applied to two common messaging use cases (\textit{Account Driven} use case and \textit{Data Distribution} use case) and examined whether they have significance difference in Power Consumption.  
We concluded that the comparison between the strategies mentioned did not have significant difference on a small scale, however at large scale and long term usage, where economy of scale is important, this difference can have a significant energy impact and thus lead to energy savings or waste. 
For the future, we anticipate to perform a deeper analysis of the data, in order to create a pricing model for messaging systems that takes into account energy consumption. Furthermore, relevant performance models can be created that guide towards optimized configurations for a needed setup.

As a more general remark, in order to better enhance sustainability in large scale services like Cloud Computing, a similar approach could be applied to other types of services or architectures, leading to a user-level or action-level accountability on the energy usage of operations. This, if applied at a policy level, could motivate system engineers and architects to investigate such issues a priori and lead to a more sustainable software and system development.

\bibliographystyle{abbrv}
\bibliography{bibliography}

\begin{thebibliography}{10}

\bibitem{rabbitmqAMQP091}
{A}{M}{Q}{P} 0-9-1 {S}pecification | {R}abbit{M}{Q} --- rabbitmq.com.
\newblock \url{https://www.rabbitmq.com/tutorials/amqp-concepts#exchanges}.
\newblock [Accessed 17-12-2024].

\bibitem{dshb}
{G}it{H}ub - beltex/dshb: mac{O}{S} system monitor --- github.com.
\newblock \url{https://github.com/beltex/dshb}.
\newblock [Accessed 11-09-2024].

\bibitem{powerstat}
{G}it{H}ub - {C}olin{I}an{K}ing/powerstat: {P}owerstat - power consumption measurement using stats {I}ntel {R}{A}{P}{L} interface --- github.com.
\newblock \url{https://github.com/ColinIanKing/powerstat}.
\newblock [Accessed 11-09-2024].

\bibitem{pika}
{G}it{H}ub - pika/pika: {P}ure {P}ython {R}abbit{M}{Q}/{A}{M}{Q}{P} 0-9-1 client library --- github.com.
\newblock \url{https://github.com/pika/pika}.
\newblock [Accessed 13-12-2024].

\bibitem{energytracker}
{G}it{H}ub - rdegges/energy-tracker --- github.com.
\newblock \url{https://github.com/rdegges/energy-tracker}.
\newblock [Accessed 11-09-2024].

\bibitem{grafana}
{G}rafana: {T}he open observability platform | {G}rafana {L}abs --- grafana.com.
\newblock \url{https://grafana.com/}.
\newblock [Accessed 11-09-2024].

\bibitem{githubVoreakou}
{G}reen {S}ystem {D}esign - {S}etup {G}uide for ubuntu os.
\newblock \url{https://github.com/MariaVoreakou/Green-System-Design }.
\newblock [Accessed 06-12-2024].

\bibitem{intelHyperThreading}
{I}ntel {H}yper-{T}hreading --- intel.com.
\newblock \url{https://www.intel.com/content/www/us/en/gaming/resources/hyper-threading.html}.
\newblock [Accessed 22-01-2025].

\bibitem{intelTurboboost}
{I}ntel® {T}urbo {B}oost {T}echnology --- intel.com.
\newblock \url{https://www.intel.com/content/www/us/en/gaming/resources/turbo-boost.html}.
\newblock [Accessed 22-01-2025].

\bibitem{kepler}
{K}epler as {R}{P}{M} --- sustainable-computing.io.
\newblock \url{https://sustainable-computing.io/installation/kepler-rpm/}.
\newblock [Accessed 11-09-2024].

\bibitem{locust}
{L}ocust.io --- locust.io.
\newblock \url{https://locust.io/}.
\newblock [Accessed 11-09-2024].

\bibitem{nvidianvml}
{N}{V}{I}{D}{I}{A} {M}anagement {L}ibrary ({N}{V}{M}{L}) --- developer.nvidia.com.
\newblock \url{https://developer.nvidia.com/management-library-nvml}.
\newblock [Accessed 22-01-2025].

\bibitem{hwmonitor}
{O}pen {H}ardware {M}onitor --- openhardwaremonitor.org.
\newblock \url{https://openhardwaremonitor.org/}.
\newblock [Accessed 11-09-2024].

\bibitem{openmessaging}
{O}pen{M}essaging --- openmessaging.cloud.
\newblock \url{https://openmessaging.cloud/}.
\newblock [Accessed 17-01-2025].

\bibitem{profiles}
{P}rofiles :: {S}pring {B}oot --- docs.spring.io.
\newblock \url{https://docs.spring.io/spring-boot/reference/features/profiles.html}.
\newblock [Accessed 11-09-2024].

\bibitem{rabbitmq}
{R}abbit{M}{Q} {D}ocumentation | {R}abbit{M}{Q} --- rabbitmq.com.
\newblock \url{https://www.rabbitmq.com/docs/use-rabbitmq}.
\newblock [Accessed 11-09-2024].

\bibitem{scaphandre}
{S}caphandre {D}ocumentation --- hubblo-org.github.io.
\newblock \url{https://hubblo-org.github.io/scaphandre-documentation/index.html}.
\newblock [Accessed 11-09-2024].

\bibitem{GeTh}
A.~Bagios.
\newblock {G}it{L}ab -- {G}eneralized {T}ransmission {H}ub.
\newblock \url{https://gitlab.com/hua-dev/geth}.
\newblock [Accessed 11-09-2024].

\bibitem{bagios}
A.~Bagios.
\newblock Unified iot querying system.
\newblock Master's thesis, Department Informatics and Telematics, Harokopio University of Athens, 2024.
\newblock Available at \url{https://estia.hua.gr/file/lib/default/data/29056/theFile}.

\bibitem{energyImpact}
R.~Buyya, S.~Ilager, and P.~Arroba.
\newblock Energy-efficiency and sustainability in new generation cloud computing: A vision and directions for integrated management of data centre resources and workloads.
\newblock {\em Software: Practice and Experience}, 54(1):24--38, 2024.

\bibitem{cadorel2024protocol}
E.~Cadorel and D.~Saingre.
\newblock A protocol to assess the accuracy of process-level power models.
\newblock In {\em 2024 IEEE International Conference on Cluster Computing (CLUSTER)}, pages 74--84. IEEE, 2024.

\bibitem{powerapi}
G.~Fieni, D.~R. Acero, P.~Rust, and R.~Rouvoy.
\newblock Powerapi: A python framework for building software-defined power meters.
\newblock {\em Journal of Open Source Software}, 9(98):6670, 2024.

\bibitem{RANScaphandre}
V.~Gudepu, R.~R. Tella, C.~Centofanti, J.~Santos, A.~Marotta, and K.~Kondepu.
\newblock Demonstrating the energy consumption of radio access networks in container clouds.
\newblock In {\em NOMS2024, the IEEE/IFIP Network Operations and Management Symposium}, 2024.

\bibitem{decentralisedmqttPatients}
K.~N. Haque, J.~Islam, I.~Ahmad, and E.~Harjula.
\newblock Decentralized pub/sub architecture for real-time remote patient monitoring: A feasibility study.
\newblock In {\em Nordic Conference on Digital Health and Wireless Solutions}, pages 48--65. Springer, 2024.

\bibitem{comparisoncpugpu}
M.~Jay, V.~Ostapenco, L.~Lef{\`e}vre, D.~Trystram, A.-C. Orgerie, and B.~Fichel.
\newblock An experimental comparison of software-based power meters: focus on cpu and gpu.
\newblock In {\em 2023 IEEE/ACM 23rd International Symposium on Cluster, Cloud and Internet Computing (CCGrid)}, pages 106--118. IEEE, 2023.

\bibitem{rapl}
K.~N. Khan, M.~Hirki, T.~Niemi, J.~K. Nurminen, and Z.~Ou.
\newblock Rapl in action: Experiences in using rapl for power measurements.
\newblock {\em ACM Transactions on Modeling and Performance Evaluation of Computing Systems (TOMPECS)}, 3(2):1--26, 2018.

\bibitem{lin2024bridging}
C.~Lin and M.~Shahrad.
\newblock Bridging the sustainability gap in serverless through observability and carbon-aware pricing.
\newblock In {\em 3rd Workshop on Sustainable Computer Systems (HotCarbon’24)}, page~41, 2024.

\bibitem{prometheus}
Prometheus.
\newblock {M}onitoring system \& time series database --- prometheus.io.
\newblock \url{https://prometheus.io/}.
\newblock [Accessed 09-01-2025].

\bibitem{greenComputing}
L.-D. Radu.
\newblock Green cloud computing: A literature survey.
\newblock {\em Symmetry}, 9(12), 2017.

\bibitem{metrics}
V.~D. Reddy, B.~Setz, G.~S.~V. Rao, G.~Gangadharan, and M.~Aiello.
\newblock Metrics for sustainable data centers.
\newblock {\em IEEE Transactions on Sustainable Computing}, 2(3):290--303, 2017.

\bibitem{microsoftQueueBasedLoad}
RobBagby.
\newblock {Q}ueue-{B}ased {L}oad {L}eveling pattern - {A}zure {A}rchitecture {C}enter --- learn.microsoft.com.
\newblock \url{https://learn.microsoft.com/en-us/azure/architecture/patterns/queue-based-load-leveling}.
\newblock [Accessed 13-12-2024].

\bibitem{socialNetworks}
E.~Sahafizadeh and B.~Tork~Ladani.
\newblock A model for social communication network in mobile instant messaging systems.
\newblock {\em IEEE Transactions on Computational Social Systems}, 7(1):68--83, 2020.

\bibitem{strempel2021measuring}
T.~Strempel.
\newblock Measuring the energy consumption of software written in c on x86-64 processors.
\newblock 2021.

\bibitem{thingsboard}
thingsboard.
\newblock {T}hings{B}oard — {O}pen-source {I}o{T}{P}latform --- thingsboard.io.
\newblock \url{https://thingsboard.io/}.
\newblock [Accessed 14-12-2024].

\end{thebibliography}



\ifCLASSOPTIONcaptionsoff
  \newpage
\fi

%








\end{document}